\documentclass[a4paper,fleqn]{cas-dc}

\usepackage[numbers]{natbib}

\usepackage{color}

\newcommand{\red}{\color{red}}

\def\tsc#1{\csdef{#1}{\textsc{\lowercase{#1}}\xspace}}
\tsc{WGM}
\tsc{QE}

\begin{document}
\let\WriteBookmarks\relax
\def\floatpagepagefraction{1}
\def\textpagefraction{.001}

\shorttitle{Floquet valley Hall solitons}    

\shortauthors{Ivanov, Kartashov}

\title [mode = title]{Floquet valley Hall edge solitons}  



%

\author[1]{Sergey K. Ivanov}[orcid=0000-0001-9230-1035]

\cortext[1]{Corresponding author}



\ead{sergei.ivanov@uv.eu}



\affiliation[1]{organization={ICFO-Institut de Ciencies Fotoniques, The Barcelona Institute of Science and Technology},
            addressline={Av. Carl Friedrich Gauss, 3}, 
            postcode={08860},
            city={Castelldefels (Barcelona)},
            country={Spain}}



\author[2]{Yaroslav V. Kartashov}[orcid=0000-0001-8692-982X]




\affiliation[2]{organization={Institute of Spectroscopy, Russian Academy of Sciences},
            addressline={Fizicheskaya Str., 5}, 
            city={Troitsk, Moscow},
            postcode={108840}, 
            country={Russia}}












\begin{abstract}
We introduce traveling Floquet valley Hall edge solitons in a {{genuinely continuous system}}
consisting of a waveguide array with a dynamically varying domain wall between two honeycomb structures exhibiting broken inversion symmetry.
Inversion symmetry in our system is broken due to periodic and out-of-phase longitudinal modulation of the refractive index applied to the constituent sublattices of the honeycomb structure. By combining two honeycomb arrays with different initial phases of refractive index modulation we create a dynamically changing domain wall that supports localized linear Floquet edge states despite the fact that on average two sublattices in each honeycomb structure forming the domain wall have the same refractive index. In the presence of focusing nonlinearity, bright or dark Floquet edge solitons may bifurcate from such linear Floquet edge states.
{{We numerically identified family of these solitons and compared them with the results obtained from the analytical approach, which involved averaging over one longitudinal period in the evolution coordinate.}}
These solitons exhibit localization in both spatial directions - along the interface due to nonlinear self-action and across the interface as the edge states - that allows them to travel along the domain wall over long distances without noticeable shape variations.
\end{abstract}











\maketitle


Edge states with controllable localization properties that may form at the interfaces of a broad range of topological and non-topological structured materials in acoustics, photonics, solid-state physics, and optoelectronics offer unique opportunities for control of surface currents, their routing and protection in specially engineered materials, even when the latter are insulating in the bulk. Photonic systems are particularly interesting for the realization of ultrafast switches, routers, and couplers based on the edge states, as in many cases such systems allow to dynamically tune the properties of the edge states, see~\cite{Lu-14, Ozawa-19} for recent reviews. Among the promising strategies for achieving tunable devices based on the edge states is the utilization of the material nonlinearity that can be particularly strong in certain photonic systems \cite{Smirnova-20, Bardyn-16, Dobrykh-18}. This advantage has spurred research into nonlinear edge states, encompassing both topological and nontopological variants, in various photonic structures, such as optically-induced or laser-written waveguide arrays, photonic crystals, or structured microcavities. Notably, investigations were focused on the analysis of such nonlinear effects as modulational instability of the edge states~\cite{instab01, KartashovSkryabin-16}, formation of edge solitons in periodic materials~\cite{edgesol01, edgesol03, AblowitzCurtis-14, ZhangWang-2020}, optical isolation~\cite{ZhouWang-17}, and bistability of the edge states~\cite{KartashovSkryabin-17, bistability01}, among others.

Particular attention has been dedicated to study of nonlinear edge states in systems with parameters periodically varying in evolution variable (time or propagation distance). The dynamic modulation of parameters in periodic optical systems can significantly alter their modal spectra, giving rise to intriguing phenomena such as dynamic localization, diffraction management, stimulated mode conversions, and the emergence of unconventional localized states~\cite{dinlattices01}. When this modulation is sufficiently fast (i.e., not adiabatic), it enables Floquet engineering of band structures of periodic materials, a concept crucial not only in optics, but also in quantum mechanics~\cite{floquet04, floquet02}. Floquet systems provide unprecedented opportunities for creation of artificial gauge fields for electromagnetic waves~\cite{Lu-14,Ozawa-19}, achieving topologically nontrivial phases, generation of unidirectional edge states, as proposed in~\cite{floquet05, floquet06}, and experimentally observed at optical frequencies in modulated waveguide arrays~\cite{floquet07, floquet08, floquet09}, their coupling and propagation velocity control~\cite{control01, control02}, and generation of periodically oscillating states of topological origin~\cite{pimode01, pimode02, pimode03, pimode04, pimode05}. Nonlinear self-action in such Floquet systems enable the formation of topological edge solitons.
{{These solitons are unique states that exhibit topological protection, appearing in various shapes and featuring unusual interactions. For instance, Floquet edge solitons in helical waveguide arrays have been theoretically predicted in continuous systems as strongly localized states~\cite{edgesol03} and envelope solitons~\cite{edgesol05}, as well as in discrete~\cite{edgesol04} and Dirac~\cite{edgesol09} models. Additionally, Bragg solitons~\cite{edgesol08}, multicomponent solitons~\cite{edgesol06}, and solitons in media with non-Kerr-type nonlinearity~\cite{edgesol10} have been studied.
Topological optical solitons have been observed experimentally~\cite{solobsv01, solobsv02, ArkhipovaZhang-23}}}
and nonlinearity-induced topological Floquet insulators were reported too~\cite{solobsv03}.

A different class of physical systems also admitting the formation of traveling edge states, where the interplay between nonlinearity and light localization at the edge was considered, is represented by valley Hall systems~\cite{Dong-17,Gao-17,Wu-17,Noh-18,Bleu-18,He-19,Shalaev-19,Zeng-20,Zhong-20,Gong-20, Liu-21, Xue-21, Xi-20}. The formation of edge states in valley-Hall systems is linked to the disruption of the inversion symmetry, rather than due to breakup of the time-reversal symmetry, as it occurs in some of the Floquet systems mentioned above. Nonlinearity in valley Hall systems may also cause modulational instabilities of the edge states and formation of valley Hall solitons, see~\cite{Smirnova-21, Zhong-21}. Different types of such states have been discovered, including scalar bright and dark~\cite{Tang-21,TangRen-22,TangBelic-22}, vector~\cite{TangZhangKartashov-22} and Dirac edge solitons~\cite{TianZhangLi-22, TianWang-23}.


These two classes of physical systems allowing observation of intriguing dynamics of the travelling edge states, were nevertheless considered separately up to now due to different mechanisms behind edge state formation in them. The goal of this work is to introduce a combined Floquet valley Hall system, inheriting representative properties from both classes, including the possibility of formation of the edge states at the dynamic interface between two honeycomb structures that on average are identical. Namely, we consider the interface between two honeycomb structures with periodically oscillating refractive index modulation depths in two constituent sublattices, with phase of this modulation being different at two sides of the domain wall. We find that localized in-gap Floquet modes emerge at the interface of two arrays, while the spectrum of this system becomes periodic along the quasi-propagation constant axis, a representative feature of Floquet systems. In a focusing medium, these oscillating states play a pivotal role in generation of bright and dark Floquet valley Hall solitons.
{{We demonstrate that for small deviation of quasi-propagation constant of solitons from quasi-propagation constant corresponding to linear edge modes, the dynamics of solitons can be accurately described by the nonlinear Schr\"{o}dinger equation for the mode envelopes, obtained by averaging over one longitudinal period. Additionally, for large deviations of quasi-propagation constant from linear value, when the soliton becomes narrow, we found soliton solutions numerically. These solitons are stable and can propagate undistorted over very long distances despite their strong localization.}}


\begin{figure*}[t!]
\centering
\includegraphics[width=1.0\linewidth]{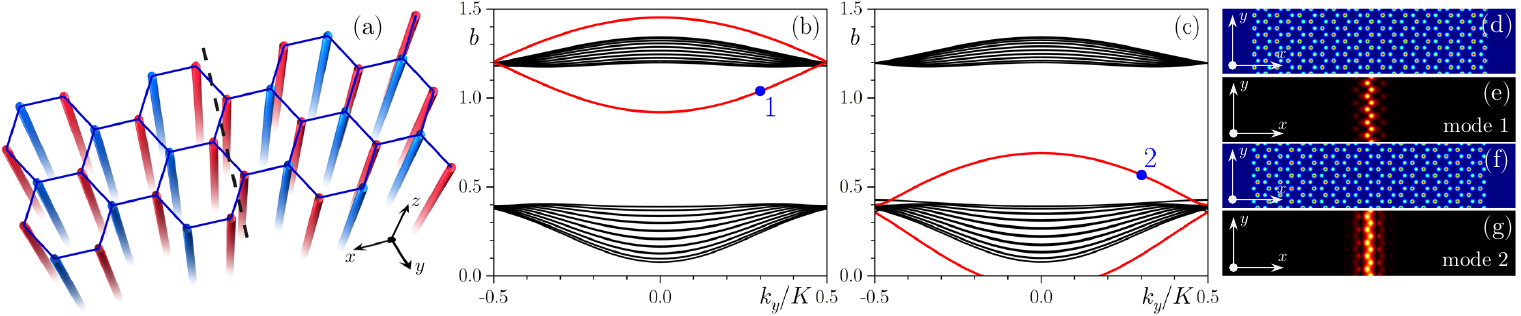}
\caption{
(a) Schematic representation of static honeycomb structure with a domain wall between two arrays with opposite detuning of two sublattices. Two sublattices are distinguished by different colors. Domain wall between two honeycomb arrays with different detunings is highlighted by the black dashed line. Band structures of the arrays with the valley Hall domain wall for $p_1=4$ and $p_2=6$ (b) and $p_1=6$ and $p_2=4$ (c). The black curves correspond to bulk states, while the red curves correspond to the valley Hall edge states. Profile of static honeycomb waveguide array (d) and profile of the valley Hall edge state with $k=0.3K$ (e) for $p_1=4$ and $p_2=6$. Array (f) and profile of the edge state with $k=0.3K$ (g) for $p_1=6$ and $p_2=4$. Edge states in (e),(g) correspond to the blue dots in panels (b),(c).
{{The arrays and states are shown within the window $-39 \leq x \leq 39$ and $-17.3 \leq y \leq 17.3$.}}
Here and below $d=2$ and $\sigma=0.5$.
}
\label{fig1}
\end{figure*}

We consider the paraxial propagation of light along the $z$-axis in a modulated waveguide array with focusing nonlinearity. The propagation dynamics can be described by the dimensionless nonlinear Schr\"{o}dinger equation for field amplitude $\psi$:
\begin{equation} 
\label{NLS}
    i \frac{\partial \psi}{\partial z} =  -\frac{1}{2} \nabla^2 \psi  - \mathcal{R}(x,y,z)\psi - |\psi|^2\psi.
\end{equation}
Here $\nabla=(\partial/\partial x,\partial/\partial y)$, the $x$ and $y$ coordinates are normalized to the characteristic transverse scale $r_0=10$~$\mu \textrm{m}$, the propagation distance $z$ is normalized to the diffraction length {{$kr_0^2\approx 1.14 \, \text{mm}$}}, $k = 2\pi n_0/\lambda$ is the wavenumber at working wavelength $\lambda=800$~$\textrm{nm}$, and $n_0\approx 1.45$ is the unperturbed refractive index of the material (for example, fused silica). Dimensionless intensity $|\psi|^2=1$ corresponds to peak intensity of $I=n_0|\psi|^2/k^2r_0^2n_2\approx 8\cdot10^{15} \, \textrm{W}/\textrm{m}^2$, where the nonlinear refractive index is $n_2\sim1.4\cdot10^{-20} \, \textrm{m}^2/\textrm{W}$. Here we consider honeycomb waveguide arrays consisting of two sublattices with indices $1$ and $2$, so that the function $\mathcal{R}(x,y,z) = p_1(z)\sum_{n,m} Q_{n,m}^{(1)}(x,y)+ p_2(z)\sum_{n,m} Q_{n,m}^{(2)}(x,y)$ describes refractive index distribution inside the array constructed from single-mode Gaussian waveguides, where profiles of the individual waveguides are described by the function $Q_{n,m}^{(i)}(x,y) = e^{-[(x-x_{n,m}^{(i)})^2+(y-y_{n,m}^{(i)})^2]/\sigma^2}$, while $x_{n,m}^{(i)},y_{n,m}^{(i)}$ represent coordinates of the nodes of each sublattice with index $i=1,2$. Each waveguide has a width of $\sigma = 0.5$. The depth of sublattices is given by $p_{1,2}(z) = k^2r_0^2\delta n_{1,2}(z)/n_0$, where $\delta n_{1,2}(z)$ are $z$-modulated refractive index contrasts. We consider a configuration that is periodic along the $y$-axis and is limited along the $x$-axis with outer boundaries located far away from the domain wall. Thus, $R(x,y,z) = R(x,y+L,z)=R(x,y,z+Z)$, where $Z$ is the longitudinal period of the refractive index modulation, $y$-period is given by $L=3^{1/2}d$, and $d=2$ is the spacing between neighbouring waveguides.
{{This corresponds to a $20$-$\mu\text{m}$ distance between waveguides, a waveguide width of $5 \, \mu\text{m}$ and lattice $y$-period of about $34.6 \, \mu\text{m}$; while the propagation distance $z=1$ corresponds to about $1.14\,\text{mm}$.}}
Notice that the domain wall in valley Hall system can be created by stacking two arrays, where sublattice depths $p_1$ and $p_2$ are swapped [see a typical configuration of the domain wall in static unmodulated valley Hall structure in Fig. \ref{fig1}(a)].


\begin{figure*}[t!]
\centering
\includegraphics[width=\linewidth]{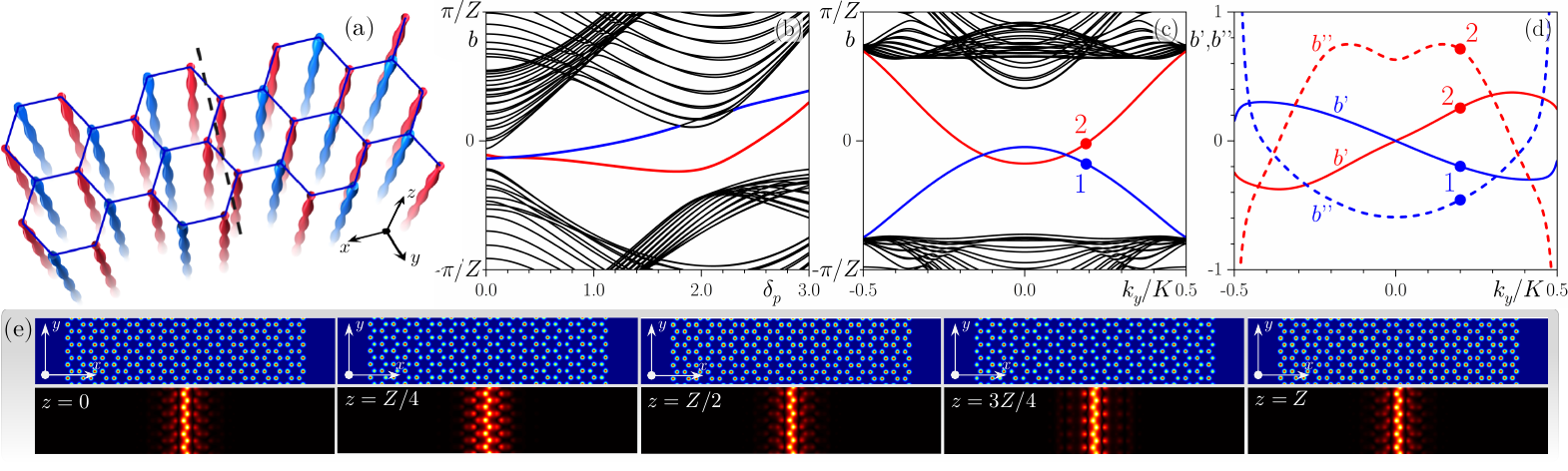}
\caption{
(a) Schematic representation of Floquet array with dynamic domain wall. Two sublattices with $p_1(z)$ and $p_2(z)$ are distinguished by different colors. Domain wall between two honeycomb arrays with different detunings is highlighted by the black dashed line. (b) Floquet spectrum of the modulated array with $p_{{\rm av}}=5$ at $k_y=0.0K$ versus detuning $\delta_p$ and (c) Floquet spectrum vs Bloch momentum $k_y$ for modulated array with $p_{{\rm av}}=5$ and $\delta_p=1$. The black curves are the bulk states, while the blue and red curves are the Floquet valley Hall edge states. (d) First-order $b^\prime$ (solid curves) and second-order $b^{\prime \prime}$ (dashed curves) derivatives of the propagation constant of the Floquet valley Hall edge state. The blue and red dots depicted in panels (c) and (d) correspond to the modes used for construction of the bright and dark solitons, respectively. (e) Profiles of the Floquet array and profiles of Floquet valley Hall edge states with $k_y=0.2K$ at different distances $z$. This edge state corresponds to the blue dots in panels (c) and (d). Here and below longitudinal modulation period is $Z=12$.
}
\label{fig2}
\end{figure*} 

Linear eigenmodes of the $z$-modulated system that is also periodic along the $y$-direction are Floquet-Bloch waves $\psi(x,y, z) = \phi(x,y, z) e^{ibz} = u(x,y,z) e^{ibz + ik_y y}$. Here, $\phi$ exhibits periodicity along the $z$-axis: $\phi(x,y,z)=\phi(x,y,z+Z)$, while $u$ demonstrates periodicity along both the $y$ and $z$ axes: $u(x,y,z)=u(x,y+L,z)=u(x,y,z+Z)$. The Bloch momentum $k_y$ varies within the transverse Brillouin zone of width $K=2\pi/L$. Here, quasi-propagation constant $b \in [-\pi/Z, +\pi/Z)$
{{, defined modulo $\omega=2\pi/Z$,}}
represents a $k_y$-dependent eigenvalue of the problem $[i\partial/\partial z + (1/2)\nabla^2 + R(x,y,z)]\phi = b\phi$~\cite{floquet06}. Modes of the system are localized along the $x$-axis, with $u$ and $\phi$ approaching zero as $x\rightarrow \pm\infty$.

We start our analysis by considering static valley Hall domain wall between two honeycomb arrays with depths of sublattices given by $p_1 = p_{\rm av} - \delta_p$ and $p_2 = p_{\rm av} + \delta_p$, where $p_{\rm av}$ represents the average value of the waveguide depth, while $\delta_p$ denotes the detuning. We set $p_{\rm av}=5$ which corresponds to $\delta n\approx5.6\cdot 10^{-4}$. To create the domain wall, we connect two honeycomb arrays with opposite signs of sublattice detuning $\delta_p$ in the regions $x<0$ and $x>0$ [see Fig.~\ref{fig1}(a) for the schematic representation of the structure with a domain wall indicated by the black dashed line]. In this static (i.e. $z$-independent) structure $\phi$ is $z$-independent and the eigenvalue $b$ becomes conventional propagation constant. Using plane-wave expansion method, we calculated the linear projected spectrum $b(k_y)$ of the array with domain wall for $\delta_p=1$. The array profile for this case, when only deep waveguides are located on the domain wall, is shown in Fig.~\ref{fig1}(d) {{within the window $-39 \leq x \leq 39$ and $-17.3 \leq y \leq 17.3$. The unit cell of this structure is represented by a stripe that is periodic in the vertical direction, with periodicity determined by the lattice constant $L$. It extends across many unit cells in the horizontal direction and includes the domain wall.}} Two bands of this array (black lines) are depicted in Fig.~\ref{fig1}(b), revealing the formation of a wide gap between them. Remarkably, the array with inverted detuning $\delta_p = -1$, where only shallow waveguides are located on the domain wall [see Fig.~\ref{fig1}(f)], exhibits identical band structure, as showcased in Fig.~\ref{fig1}(c). However, these two cases differ by the location of the edge states corresponding to the red lines: for positive detuning they appear around upper band {(because in this case the edge states are mainly localized on deeper waveguides belonging to the domain wall and on this reason they have increased propagation constants)}, while for negative detuning they appear near the lower band {(because in this case edge states form on the domain wall containing shallower waveguides, resulting in reduced propagation constants), see also \cite{Noh-18}}. In each case, one observes two distinct types of localized along the $x$-axis edge states: one of them resides within the topological gap, while the other one appears in the trivial gap. {These two edge states for a given sign of detuning $\delta_p$ differ by the phase structure of the $u(x,y)$ function [$y$-periodic part of the Bloch wave $\psi(x,y,z) =  u(x,y) e^{ibz + ik_y y}$]. Thus, for positive detuning $\delta_p = +1$, we observe that adjacent spots in the $u(x,y)$ function along the domain wall are in-phase for state residing in the upper trivial gap, while they are out-of-phase in state residing in the topological gap between two bands. Conversely, for negative detuning $\delta_p = -1$, the function $u(x,y)$ from the topological gap consists of in-phase spots along the domain wall, while its counterpart from trivial gap laying below topological one displays out-of-phase spots. Furthermore, localization of these modes along the $x$-axis depends on the momentum $k_y$ and decreases when the mode approaches the bulk band shown by black lines.} Two representative $|\psi(x,y)|$ distributions of edge states residing in topological gap, marked by the blue dots on red branches at $k_y = 0.3K$, are shown in Figs.~\ref{fig1}(e) and \ref{fig1}(g). A typical feature of the valley Hall system is that the edge state does not connect two bands, but emanates from and disappears in the same band.

{{It is important to emphasize that although a forbidden gap emerges in the band structure, the Berry curvature $\Omega$ of the first and second bulk bands satisfies the condition $\Omega(-k) = -\Omega(k)$. This indicates that standard Chern numbers (calculated over the entire Brillouin zone) of the two upper bulk bands remain zero~\cite{Xiao-10} in valley Hall system. In this case, six valleys appear in the spectrum, with three valleys around the $K$($K^\prime$) points being equivalent. The valley Chern number of a specific valley is determined to be either $+1/2$ or $-1/2$. If the valley Chern number for a certain valley is $+1/2$ for a lattice with $\delta_p < 0$, then for a lattice with $\delta_p > 0$, the Chern number for the same valley will be $-1/2$~\cite{Noh-18}. Therefore, by designing a composite honeycomb lattice with a domain wall (highlighted by the dashed line in Fig.~\ref{fig1}(a)) between two inversion-symmetry-broken honeycomb lattices with opposite valley detunings, the Chern numbers on both sides of the interface become opposite. In this scenario, the bulk-edge correspondence principle, applied to a valley Hall system, predicts the formation of edge states localized at the domain wall, decaying perpendicularly to it. The emergence of such edge states is a manifestation of the valley Hall effect~\cite{Liu-21,Xue-21}.}}

Next, we turn our attention to a $z$-modulated waveguide array in linear regime, where waveguide depths $p_1(z) = p_{\rm av} - \delta_p \sin(2\pi z/Z)$ and $p_2(z) = p_{\rm av} + \delta_p \sin(2\pi z/Z)$ in two sublattices periodically change with distance, as illustrated in Fig.~\ref{fig2}(a). In this figure waveguide depth variations is schematically represented by the oscillations of red and blue channels belonging to different sublattices. Each honeycomb array forming the domain wall spends now half of the longitudinal period $Z$ in one valley Hall ``phase'' with $p_1<p_{\rm av}<p_2$ and other half-period in other ``phase'' with $p_1>p_{\rm av}>p_2$.
{{Remarkably, on average two honeycomb structures forming the interface are identical, and the interface is actually created by the difference in phase of the longitudinal refractive index variation in two sublattices in two arrays (in each of them depths of two sublattices oscillate out-of-phase).}} A somewhat related idea of localization control in ``line'' helical waveguide arrays has been proposed in \cite{phaseloc}. We found that despite notable longitudinal refractive index variations in two arrays, the domain wall between them still supports $z$-periodic Floquet edge states that demonstrate distinguishing features typical for both valley Hall and Floquet systems.

{{Fig.~\ref{fig2}(b) shows the first Brillouin zone $b \in [-\pi/Z, +\pi/Z)$ as a function of $\delta_p$, obtained for $k_y=0K$ and $Z=12$ which in the physical units corresponds to the modulation period of $13.64\,\text{mm}$.}} {{To calculate these Floquet band structures we first obtain Bloch modes $u^{st}_i(x,y)$ originating from the two uppermost bands of the truncated array with static valley Hall domain wall, utilizing a plane-wave expansion technique. The number of such modes is $n$. Each mode, $u^{st}_i$, is normalized as $\iint_S \, {u^{st}_i}^* u^{st}_j \, d\!xd\!y=\delta_{ij}$, where the Hermitian product implies integration over a single unit cell of the array $S$. These modes were then propagated through a modulated array for one period $Z$. Longitudinal refractive index modulation induces coupling among the modes from the first two bands (neglecting coupling to lower bands, which remain well-separated). The resultant output distributions, $u^{out}_j$, corresponding to input $u^{st}_j$, were projected onto the initial basis of modes, $u^{st}_i$, yielding the $n\times n$ projection matrix, $H_{ij} = \iint_S \, {u^{st}_i}^* u^{out}_j \, d\!xd\!y$, whose eigenvalues represent the Floquet exponents, $\exp(i\beta)$. Quasi-propagation constants are determined as $b = \beta/Z$. Obviously, this procedure shows that quasi-propagation constants of our Floquet system are defined modulo $\omega=2\pi/Z$, i.e. spectrum shown in Fig.~\ref{fig2}(b) is periodic along the $b$ axis with period $\omega$.}} The red and blue curves in Fig.~\ref{fig2}(b) represent localized edge states in the Floquet spectrum of longitudinally modulated structure. These Floquet states emerge from the point of overlap of the bulk band (black curves) with its own Floquet replica, as a consequence of the spectrum's periodicity along the $b$ axis defined by modulation frequency $\omega$. {This overlap results in ``interaction'' of replicas of the same band (due to $z$-modulation) and may lead to opening of the gap between them. Notably, this behavior resembles the behaviour of band structures in anomalous topological insulators, where the topological phase is intricately linked to the modulation parameters governing the system (see, e.g.,~\cite{Asboth-14,DalLago-15,ChengPan-19,WuSong-21,Sidorenko-15}). In such anomalous topological systems, including for example Su-Schrieffer-Heeger lattices with periodically oscillating waveguides, edge states, also known as $\pi$-modes, emerge in the gaps opening in the points of overlap of band with its own Floquet replica due to longitudinal modulation. The physical origin of the edge states in our system is similar, but now they are encountered in completely different valley Hall configuration, where localization of the modes also depends on Bloch momentum $k_y$. In our case, the non-monotonic behavior of the width of the gap opened due to longitudinal modulation as a function of $\delta_p$ is evident: initially, the gap width increases with small $\delta_p$, then it decreases, nearly closing the gap. Subsequently, as $\delta_p$ becomes large, the gap width increases again. Because localization of the edge states usually increases with increase of the gap width, the optimal selection of the amplitude of modulation is important, hence our choice of $\delta_p \sim 1$.}

In Fig.~\ref{fig2}(c) we present quasi-propagation constants $b$ of all modes of this Floquet system versus Bloch momentum $k_y$, with red and blue lines corresponding to the states localized on the domain wall. One can clearly see that there are two edge states in the spectrum, whose existence is crucial for subsequent construction of both bright and dark edge solitons. It is noteworthy that as $\delta_p$ approaches zero, the propagation constants of the two edge states tend to merge with the bulk bands. Likewise, the edge states are most localized when their quasi-propagation constant is close to the center of the gap in Fig.~\ref{fig2}(c), and gradually delocalize when quasi-propagation constant approaches the bulk band.
{{Similarly to static systems, the Floquet edge states in dynamic systems can be distinguished by their profiles. In Fig.~\ref{fig2}(c), the blue curve corresponds to a state with an in-phase field between adjacent waveguides along the domain wall, while the red curve corresponds to a state with out-of-phase spots in adjacent waveguides.}}

To characterize the dispersion properties of the edge states, we show the first-order derivative $b^\prime = \partial b/\partial k_y$ (solid curves) and the second-order derivative $b^{\prime\prime} = \partial^2 b/\partial k_y^2$ (dashed curves) of their quasi-propagation constant in Fig.~\ref{fig2}(d). The former determines the group velocity of the edge state $v = -b^\prime$, while the latter determines its dispersion rate (see \cite{edgesol06}). Given that $b^{\prime\prime}>0$ across a wide range in the first transverse Brillouin zone for the red curve and $b^{\prime\prime}<0$ for the blue curve, it is plausible that such modulated valley Hall system may support both bright and dark Floquet valley Hall edge solitons localized on the domain wall for the same set of parameters and for the same value of Bloch momentum $k_y$. Floquet valley Hall edge states exhibit substantial shape variations during propagation while remaining precisely $Z$-periodic, as evident in Fig.~\ref{fig2}(e). The $|\psi|$ distributions of these modes at selected distances for $\delta_p = 1$ are displayed along with corresponding array profiles, revealing identical field modulus distributions at distances $z=0$, $Z/2$, and $Z$. However, the profiles at $z=Z/4$ and $z=3Z/4$ differ, underscoring that the full period corresponds to $Z$. Subsequently, we explore states with Bloch momentum $k_y = 0.2K$, emphasizing that their properties remain similar for other $k_y$ values.


\begin{figure*}[t!]
\centering
\includegraphics[width=\linewidth]{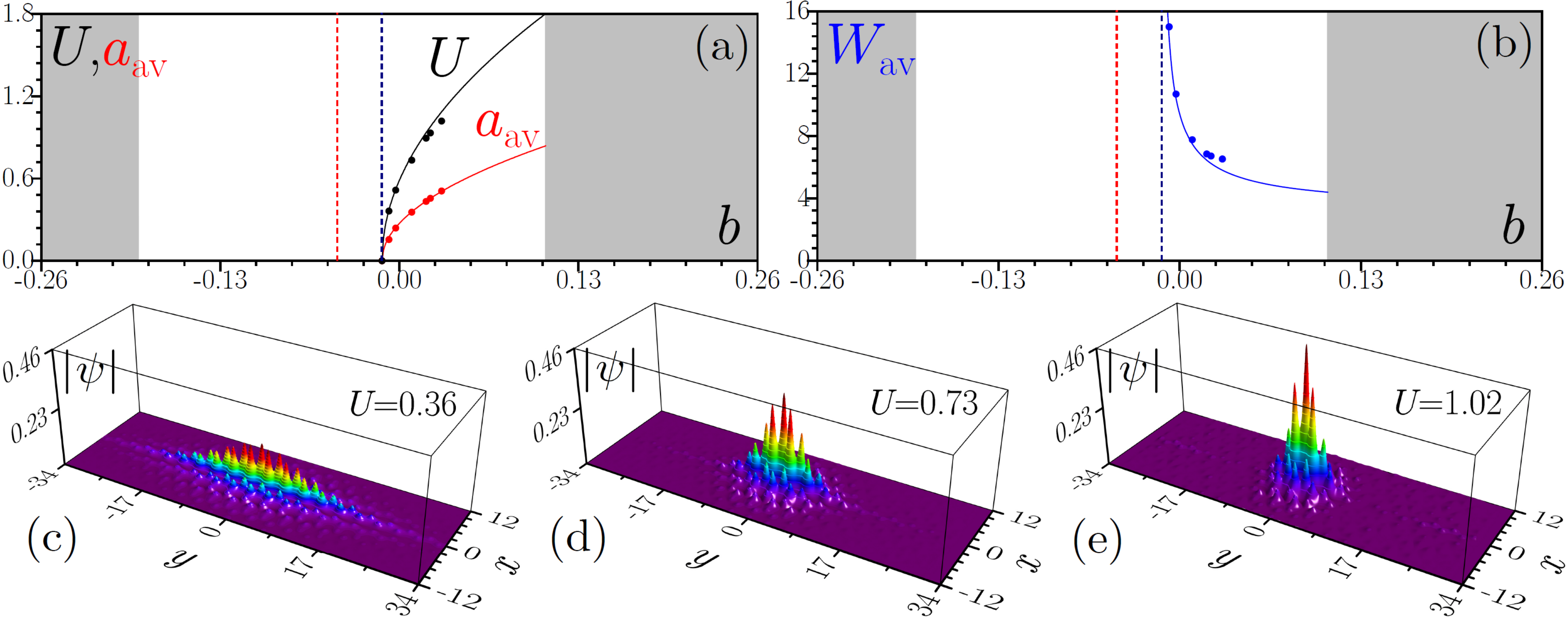}
\caption{
{{(a) Power $U$, averaged amplitude $a_{\rm av}$, and (b) averaged integral width $W_{\rm av}$ of the edge soliton at $k_y=0K$ versus quasi-propagation constant $b$. Solid lines show predictions of the envelope equation, while dots show results for exact states obtained using numerical iterations. (c)-(e) Examples of field modulus distributions in exact edge solitons with different powers $U$.}}
}
\label{fig3}
\end{figure*}

\begin{figure}[t!]
\centering
\includegraphics[width=\linewidth]{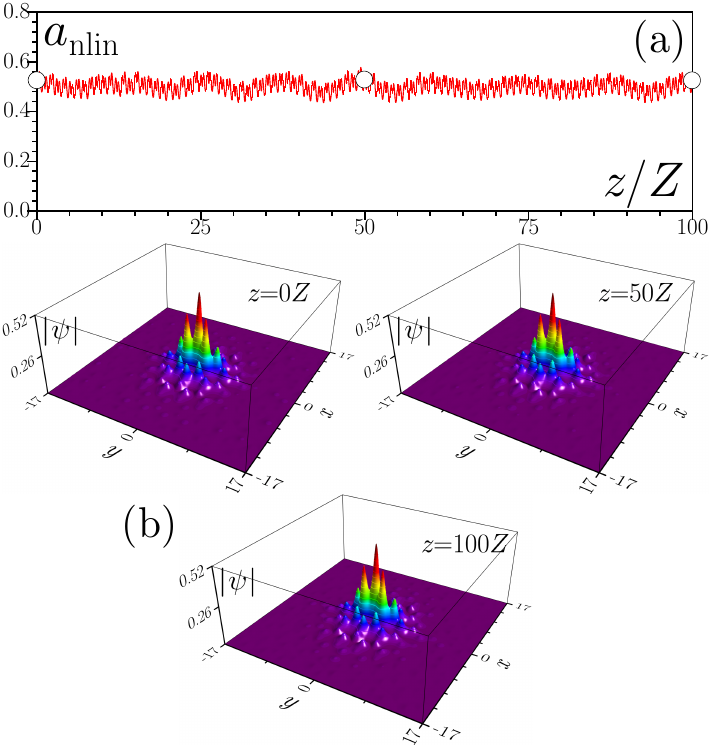}
\caption{
{{Propagation dynamics of strongly localized edge soliton with $U=1.02$ at $k_y=0K$. (a) Peak amplitude versus distance, and (b) field modulus distributions at different distances.}}
}
\label{fig4}
\end{figure}

\begin{figure*}[t!]
\centering
\includegraphics[width=\linewidth]{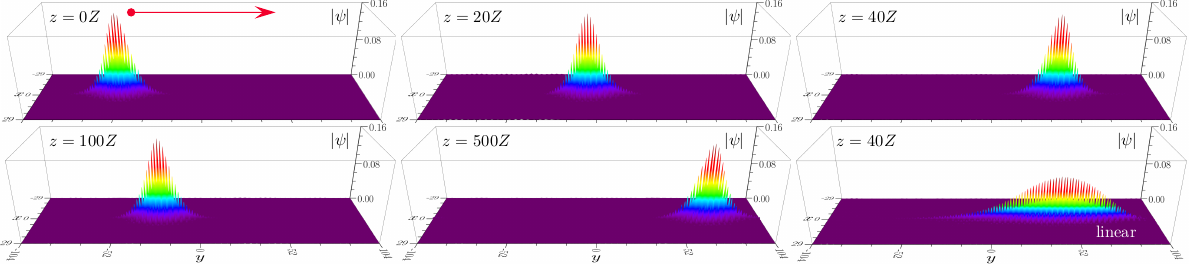}
\caption{
Propagation dynamics of the bright Floquet valley Hall edge soliton bifurcating from the blue branch in Fig.~\ref{fig2}(c) at $k_y=0.2K$ and corresponding to $b_{\rm nl}=0.005$, $v=-b^\prime=0.199$, $b^{\prime\prime}=-0.458$ and $\chi=0.132$. Red arrow indicates the direction of soliton motion. Notice that horizontal axis in all plots corresponds to $y$-axis. Right bottom panel shows profile for the same input after propagation up to $z=40Z$ without nonlinearity.
{{All amplitude distributions are shown within the window $-29 \leq x \leq 29$ and $-104 \leq y \leq 104$.}}
}
\label{fig5}
\end{figure*}

\begin{figure}[t!]
\centering
\includegraphics[width=\linewidth]{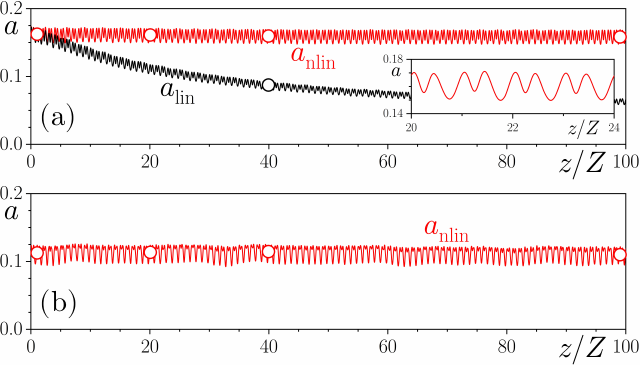}
\caption{
(a) Peak amplitude vs normalized propagation distance $z/Z$ for nonlinear $a_{\rm nlin}$ (red curve) and linear $a_{\rm lin}$ (black curve) propagation regimes of the bright envelope solution~(\ref{bright_soliton_envelope}). Red dots correspond to the field modulus distributions at different distances depicted in Fig. \ref{fig5}, while black dot corresponds to the field modulus distribution obtained for linear propagation. The inset shows evolution of the amplitude of the edge soliton over four lattice periods to stress its nearly periodic oscillations. (b) Peak amplitude vs normalized distance $z/Z$ for nonlinear $a_{\rm nlin}$ propagation of the dark envelope solution~(\ref{dark_soliton_envelope}). Red dots correspond to field modulus distributions depicted in Fig. \ref{fig8}.
}
\label{fig6}
\end{figure}

To explore the bifurcation of the family of edge solitons from linear Floquet states, we consider now Eq. (\ref{NLS}) with cubic nonlinearity included into it.
{{
We will show that for sufficiently wide solitons, the dynamics of their envelopes can be described by an effective one-dimensional nonlinear Schr\"{o}dinger equation with a spatial coordinate directed along the interface. However, first, we obtain soliton solutions numerically for the blue branch of Fig.~\ref{fig2}(c). For this, we use an iterative self-consistency method modified for $z$-periodic potentials~\cite{edgesol01}. In $z$-independent static arrays, the self-consistency algorithm involves the following steps: choosing an initial guess, finding the induced potential, calculating the eigenmodes, recalculating the induced potential through a properly weighted superposition of these modes, and iterating until convergence is reached. In our case, since the potential changes periodically in $z$, we look for solitons that are shape-preserving only periodically with the period $Z$. Nonlinear states that completely reproduce their shape are found only at $k=0K$, as only then the group velocity is zero, $v=0$. For other values of $k$, although the soliton parameters such as width and amplitude are periodic along the evolution coordinate, the states are traveling, making this method inapplicable. Using this algorithm we calculated bright solitons bifurcating from the blue branch depicted in Fig.~\ref{fig2}(c). In Fig.~\ref{fig3}(a), the dots display the relation between the quasi-propagation constant $b$ and the power $U=\iint|\psi|^2d\!xd\!y$ of the bright solitons bifurcating from corresponding linear mode (shown by vertical blue line), where the integral is computed over all lattice area, as well as relation between $b$ and the averaged peak amplitude $a_{\rm av}=Z^{-1}\int_0^Z a(z)d\!z$, where $a=\max \left(|\psi|\right)$, at $k=0K$. The figure shows that both $U$ and $a_{\rm av}$ grow monotonically but exhibit a nonlinear dependence on $b$. Fig.~\ref{fig3}(b) illustrates the corresponding total averaged width $W_{\rm av}=Z^{-1}\int_0^Z W(z)d\!z$ of the numerically found solitons, where $W=(W_x+W_y)^{1/2}$ with $W_{x}=2(U^{-1}\iint x^2 |\psi|^2d\!xd\!y)^{1/2}$ and $W_{y}=2(U^{-1}\iint y^2 |\psi|^2d\!xd\!y)^{1/2}$. Nonlinearity allows tuning of the localization of states by changing the location of $b$ in the gap. The limit as $b\rightarrow b_{\rm lin}$ corresponds to the linear limit, where $b_{\rm lin}$ represents the linear propagation constant of the state undergoing nonlinear bifurcation. An increase in the detuning from $b_{\rm lin}$ leads to a sharp decrease in the soliton width along the interface and a gradual decrease in the width across the interface. At sufficiently high $U$, when $b$ is close to the bulk band (shaded regions), the soliton couples with the bulk modes, and stable solitons were not found. Soliton profiles for various propagation constants are shown in panels of Fig.~\ref{fig3}(c)-(e). A transition from a relatively broad state extended along the domain wall [Fig.~\ref{fig3}(c)] to a highly localized soliton [Fig.~\ref{fig3}(e)] spanning only over a few adjacent waveguides is evident.

An example of the propagation of a highly localized soliton from Fig.~\ref{fig3}(e) with $U=1.02$ is shown in Fig.~\ref{fig4}. The propagation of soliton was modeled in the frames of continuous Eq.~(\ref{NLS}) accounting for all details of the modulated refractive index landscape. Here the slip-step fast Fourier method was used for modeling light propagation. As can be seen from the field modulus distributions [Fig.~\ref{fig4}(b)], this soliton covers several waveguides. The soliton survived after propagation of over one hundred longitudinal periods with its peak amplitude exhibiting an almost periodic character with a period of $Z$, remaining on average practically unchanged over this long distance [Fig.~\ref{fig4}(a)]. The soliton profiles remain consistent at different distances, and the radiation into the bulk of the array is minimal. The dimensionless peak amplitude of this soliton corresponds to $\sim 2.2\cdot 10^{15} ~\textrm{W}/\textrm{m}^2$.
}}

For the analytical approach, we employ the slowly-varying amplitude approximation, and express the nonlinear modes emerging from different linear edge state branches as $\psi\approx e^{ibz} A(Y,z)\phi$, where $A$ is the slowly varying amplitude, and $Y=y-vz$ is the coordinate in the frame moving with velocity $v = -b^\prime$, and $\phi$ is the linear edge state profile
(see~\cite{edgesol06}). The amplitude $A$ of edge soliton can be shown to satisfy the nonlinear Schr\"{o}dinger equation given by
\begin{eqnarray} \label{amplitude_equation}
    i\frac{\partial A}{\partial z} - \frac{b^{\prime\prime}}{2}\frac{\partial^2 A}{\partial Y^2} + \chi |A|^2A = 0,
\end{eqnarray}
Here, the nonlinear coefficient $\chi=Z^{-1}\int_{0}^{Z}d\!z\,\iint_S \, d\!xd\!y\,|\phi|^4$, where the integral is computed over one $y$-period of the lattice $S$, and averaging is performed over the longitudinal period $Z$. Here the field of the carrier edge state is normalized such that $\iint_S|\phi|^2d\!xd\!y=1$. Since in our case the nonlinear coefficient $\chi>0$, the equation~(\ref{amplitude_equation}) predicts the formation of either bright (if $b^{\prime\prime}<0$) or dark (if $b^{\prime\prime}>0$) envelope edge solitons. The corresponding solutions of equation~(\ref{amplitude_equation}) are given by
\begin{equation} 
\label{bright_soliton_envelope}
    A= \left(\frac{2b_{\rm nl}}{\chi}\right)^{1/2} \; \mathrm{sech}\left[ \left(-\frac{2b_{\rm nl}}{b^{\prime\prime}}\right)^{1/2}  Y \right]e^{ib_{\rm nl}z},
\end{equation}
for a bright soliton and
\begin{equation}
\label{dark_soliton_envelope}
     A= \left(\frac{b_{\rm nl}}{\chi}\right)^{1/2} \; \tanh\left[\left(\frac{b_{\rm nl}}{b^{\prime\prime}}\right)^{1/2}  Y\right]e^{ib_{\rm nl}z},
\end{equation}
for a dark soliton. Here $b_{\rm nl}$ represents the detuning of the quasi-propagation constant of soliton from quasi-propagation constant of the linear edge state, from which bifurcation occurs.
{{For the analytical solution, this nonlinearity-induced phase shift $b_{\rm nl}$ should be sufficiently small (much smaller than the longitudinal Brillouin zone) to ensure that the profile $A$ is broad and satisfy the assumption of slow variation of the soliton profile and that the total quasi-propagation constant remains in the topological gap. The detuning parameter $b_{\rm nl}$ plays a crucial role in determining both the amplitude and width of a soliton. For instance, as $b_{\rm nl}$ increases, the peak amplitude of a bright soliton monotonically rises, while its width diminishes. These characteristics are represented by the solid lines in Fig.~\ref{fig3}(a) and (b), showing good agreement with numerical simulations for small detunings $b_{\rm nl}$. In the case of dark solitons, it is evident from Eq.~(\ref{dark_soliton_envelope}) that an increase in $b_{\rm nl}$ leads to a simultaneous increase in background intensity and a reduction in width. The limit $b_{\rm nl} \to 0$ corresponds to bifurcation point from linear edge state.}}


\begin{figure*}[t!]
\centering
\includegraphics[width=\linewidth]{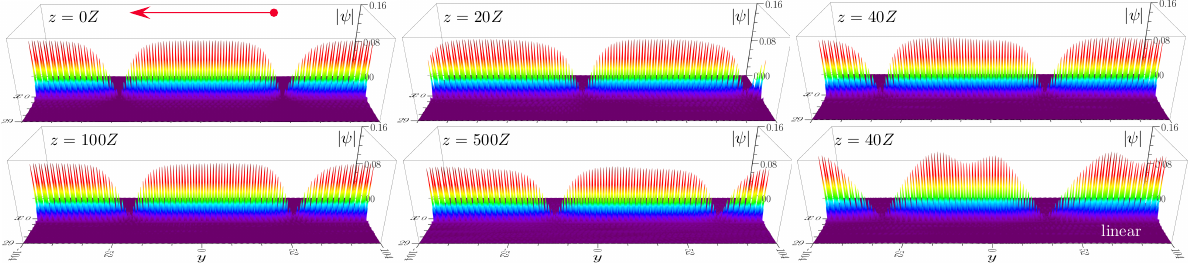}
\caption{
Propagation dynamics of the dark Floquet valley Hall edge soliton bifurcating from the red branch of Fig.~\ref{fig2}(c) at $k_y=0.2K$ and corresponding to $b_{\rm nl}=0.005$, $v=-b^\prime=-0.256$, $b^{\prime\prime}=0.711$ and $\chi=0.157$. Red arrow indicates the direction of soliton motion. Right bottom panel shows profile for the same input obtained after propagation over $z=40Z$ in linear medium. To ensure field periodicity on a large, but finite, $y$-window, a pair of well-separated dark solitons is simultaneously nested into the same edge state.
}
\label{fig7}
\end{figure*}


In Fig.~\ref{fig5} we illustrate the evolution of so-constructed bright Floquet valley Hall edge soliton with Bloch momentum $k_y=0.2K$ bifurcating from the blue branch of linear edge states in Fig.~\ref{fig2}(c). The dimensionless peak amplitude of soliton from Fig.~\ref{fig5} corresponds to $\sim 2\cdot 10^{14} ~\textrm{W}/\textrm{m}^2$. The initial condition corresponding to (obviously very efficient) excitation of this soliton solution is given by $A(y,0)\phi(x,y,0)$ with $A$ given by the equation~(\ref{bright_soliton_envelope}). Field modulus distributions in Fig.~\ref{fig5} are presented at different propagation distances illustrated in the panels. Remarkably, the soliton moves along the domain wall with minimal alterations in its shape and easily survives after five hundred longitudinal $z$-periods. The radiation into the bulk of the array is almost negligible. To affirm that these states are indeed sustained by nonlinear self-action, in Fig.~\ref{fig6}(a) we compare the evolution of the peak amplitude of this input in the nonlinear medium ($a_{\rm nlin}$, red curve) and linear medium ($a_{\rm lin}$, black curve). The peak amplitude $a_{\rm nlin}$ shows periodic oscillations around certain average level, while in the absence of nonlinearity one observes a pronounced asymmetric expansion of the wavepacket (see bottom right panel in Fig. \ref{fig5}), accompanied by a substantial decrease of its amplitude $a_{\rm lin}$.

In Fig.~\ref{fig7}, we illustrate the evolution of dark Floquet valley Hall edge soliton bifurcating from the red edge state branch of Fig. \ref{fig2}(c) at Bloch momentum $k_y=0.2K$. The initial soliton profile was constructed using the envelope $A$ defined by Eq.~(\ref{dark_soliton_envelope}). To ensure the periodicity of the field within our large, yet finite integration window spanning over $60$ $y$-periods, we simultaneously nested two well-separated dark solitons into the same edge state. The propagation of this state also demonstrates that the soliton maintains its distinctive shape while traveling along the domain wall. Remarkably, no indications of background instability are discernible for the dark soliton. The peak amplitude for this case is shown in Fig.~\ref{fig6}(b). When the nonlinearity is switched off, the background reveals considerable oscillations and two dark notches in field modulus distribution broaden significantly. This is illustrated at $z=40Z=480$ in the bottom right panel of Fig.~\ref{fig7}.

\begin{figure}[t!]
\centering
\includegraphics[width=1\linewidth]{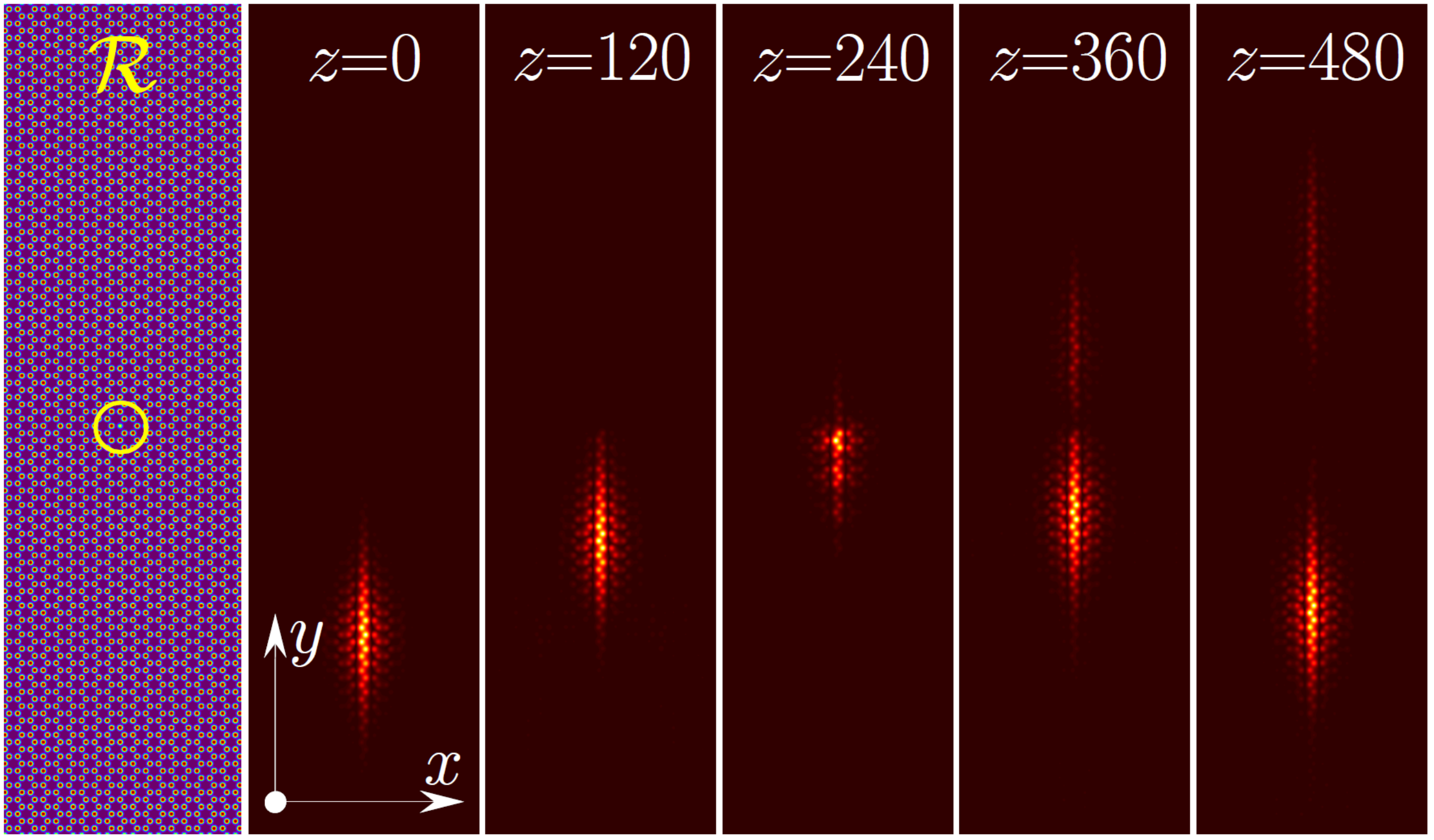}
\caption{
{
Interaction of bright soliton corresponding to $b_{\rm nl}=0.005$, $v=-b^\prime=0.199$, $b^{\prime\prime}=-0.458$, $\chi=0.132$, $k_y=0.2K$ with narrow defect at the domain wall in the form of waveguide, whose depth was reduced twice. The position of the defect is shown by the yellow circle in the panel with array profile. $|\psi|$ distributions are shown at different propagation distances.
All amplitude distributions are shown within the window $-28 \leq x \leq 28$ and $-100 \leq y \leq 100$.
}
}
\label{fig8}
\end{figure} 

{{
{We also conducted a detailed examination of the interaction between a bright Floquet edge soliton and a strong defect at the domain wall in the form of a waveguide, whose depth was halved in comparison with other waveguides (see Fig.~\ref{fig8}). Employing the same initial state as in Fig.~\ref{fig5} at $z=0$, with parameters $b_{\rm nl}=0.005$, $k_y=0.2K$, we initiated the wavepacket at $y\approx-52$, positioned far enough from the defect at $y=0$. Fig.~\ref{fig8} illustrates the array's configuration with the defect and dynamics of propagation of soliton. As shown in the figure, at the initial stage, the soliton moves in the positive direction along the $y$-axis with a group velocity of $v\approx0.2$. When the soliton collides with the defect, it is almost completely reflected. Notably, the reflected soliton retains nearly the same shape and moves with almost the same group velocity but in the opposite direction. This behavior is characteristic of topological valley Hall insulators, where counterpropagating pairs of edge states exist in the spectrum. In our case, the initial soliton at $k_y=0.2K$, interacting with a strong and narrow defect, couples with another state that has the same quasi-propagation constant but is located at $k_y=-0.2K$ and has the opposite sign of group velocity (see Fig.~\ref{fig2}). Importantly, despite the reflection, the soliton maintains its initial shape, and no scattering into the bulk of the array is observed during the collision. This demonstrates the soliton's robustness against strong defects, highlighting its protection from coupling with bulk modes.}
}}



Summarizing, we proposed a system based on the longitudinally modulated honeycomb waveguide array, where edge states simultaneously inherit distinctive features of the edge states in valley Hall and Floquet insulators. Such states appear at the domain wall separating two arrays that have the same averaged refractive index distributions, but differ by phase of modulation of the refractive index in two sublattices. Localized hybrid Floquet solitons are predicted that can propagate along such domain walls while preserving their envelopes.
{{We demonstrated two limiting types of solitons: one where the envelope spans many periods along the interface and another where the soliton is localized within a few neighboring waveguides. These solitons exhibit robustness against scattering into the bulk of the array when interacting with a defect; however, due to their valley Hall nature, they may be reflected by the defect.}}
This study introduces a new type of nonlinear Floquet states and may find applications in the design of novel photonic topological devices.
{{Moreover, our results can be generalized to other platforms where nontrivial topology can be combined with the nonlinear response of the system and applied to diverse physical systems. Indeed, the results we obtained above are generically applicable to a range of bosonic wave systems.}}

\medskip


\section*{Funding}

ICFO was supported by Grants No.~CEX2019-000910-S, No.~PID2022-138280NB-I00, and No. PRE2018-085711 (SEV-2015-0522-18-1) funded by Ministerio de Ciencia e Innovaci\'{o}n of Spain, Grant No. \\ MCIN/AEI/10.13039/501100011033/FEDER, Fundaci\'{o} Cellex, Fundaci\'{o} Mir-Puig, and Departament de Recerca i Universitats de la Generalitat de Catalunya (2021 SGR 01448) and CERCA.
Y.V.K.’s academic research was partially supported by the research project FFUU-2024-0003 of the Institute of Spectroscopy RAS and by the Russian Science Foundation (grant 24-12-00167).

\section*{Declaration of competing interest}

The authors declare that they have no known competing financial interests or personal relationships that could have appeared to
influence the work reported in this paper.

\section*{Data availability}

All data underlying the results presented in this paper may be obtained from the authors upon reasonable request.

\section*{CRediT authorship contribution statement}
All authors contribute greatly to the work.





\printcredits


\bibliography{cas-refs}

\bio{}
\endbio


\end{document}